\documentclass[aps,prb,twocolumn,groupedaddress,showpacs]{revtex4}

\usepackage{graphicx}
\begin{document}


\title{Anisotropic transport in unidirectional lateral superlattice\\
around half-filling of the second Landau level}


\author{Akira Endo}
\email[]{akrendo@issp.u-tokyo.ac.jp}
\author{Yasuhiro Iye}
\altaffiliation[Also at ]{CREST, JST Corporation.}

\affiliation{Institute for Solid State Physics, University of Tokyo\\
5-1-5 Kahsiwanoha, Kashiwa, Chiba, 277-8581 Japan}


\date{\today}

\begin{abstract}
We have observed marked transport anisotropy in short period ($a$=92 nm) unidirectional lateral superlattices around filling factors $\nu$=5/2 and 7/2: magnetoresistance shows a sharp peak for current along the modulation grating while a dip appears for current across the grating. By altering the ratio $a/l$ (with $l$=$\sqrt{\hbar/eB_\perp}$ the magnetic length) via changing the electron density $n_e$, it is shown that the $\nu$=5/2 anisotropic features appear in the range 6.6$\alt$$a/l$$\alt$7.2 varying their intensities, becoming most conspicuous at $a/l$$\simeq$6.7. The peak/dip broadens with temperature roughly preserving its height/depth up to 250 mK\@. Tilt experiments reveal that the structures are slightly enhanced by an in-plane magnetic field $B_{\|}$ perpendicular to the grating but are almost completely destroyed by $B_{\|}$ parallel to the grating. The observations suggest the stabilization of a unidirectional charge-density-wave or stripe phase by weak external periodic modulation at the second Landau level.
\end{abstract}


\maketitle

It has been known for some time that the second ($N$=1) Landau level (LL) exhibits enigmatic even denominator fractional quantum Hall effect (FQHE) \cite{Willett87,Perspectives97} at half-filling (filling factors $\nu$=5/2 and 7/2). Interestingly, the state at these filling factors is qualitatively different either from that for the lowest ($N$=0) or for the higher ($N$$\ge$2) LL's. The state at the half-filled lowest LL ($\nu$=1/2, 3/2) is now well established to be described by the Fermi sea of composite Fermions.\cite{Perspectives97} On the other hand, the ground state of higher LL's near half-filling is predicted to be a unidirectional charge-density-wave (CDW) or stripe phase by Hartree-Fock calculations,\cite{Koulakov96,Fogler96,Moessner96} which has been supported by a number of recent theories.\cite{Rezayi99,Maeda00,Shibata01} Experimentally, ultrahigh mobility ($\mu$$\agt$1000 m$^2$/Vs) two-dimensional electron gas (2DEG) at low temperatures ($T$$\alt$ 150 mK) displays strong transport anisotropy at $\nu$=9/2, 11/2, 13/2,..., between $\langle$110$\rangle$ and $\langle$1$\overline{1}$0$\rangle$ axes of the host crystal.\cite{Lilly99,Du99} It is now widely believed that the transport anisotropy is related to CDW phase, although the key factor connecting the stripe and the crystal axes still remains to be uncovered.\cite{Willett01,Cooper01}

Returning back to the second LL, $\nu$=5/2 (and 7/2) FQHE is known to be quite fragile, having a small activation energy gap, observable only for 2DEG with reasonably high mobility at very low temperatures ($T$$\alt$100 mK) \cite{Willett87,Eisenstein90,Pan99e} and to collapse in tilted magnetic fields.\cite{Eisenstein88,Eisenstein90} Moreover, it has recently been reported that an in-plane magnetic field $B_{\|}$ turns the \emph{isotropic} FQHE state into an anisotropic state similar to those observed at higher LL's, with current parallel to $B_{\|}$ giving the resistivity maxima.\cite{Pan99,Lilly99t} The role played by $B_{\|}$ in the drastic transition of the $\nu$=5/2 and 7/2 states is not exactly established yet. The most probable picture however seems to be alteration by $B_{\|}$ of Haldane's pseudopotential components.\cite{Morf98,Rezayi00} (Zeeman effect has been experimentally shown to make only small, if any, contribution.\cite{Pan01}) Whatever the role may be, the effect of $B_{\|}$ on 2DEG is generally believed to be not so large. Therefore the two states are expected to have only small energy difference. (Two-dimensional hole system exhibits anisotropic transport at $\nu$=5/2 without tilting,\cite{Shayegan00} demonstrating the subtleness of the difference.)  This in turn suggests the possibility of an alternative way to the transition.

\begin{figure}[t]
\includegraphics[bbllx=20,bblly=150,bburx=581,bbury=800,width=7.5cm]{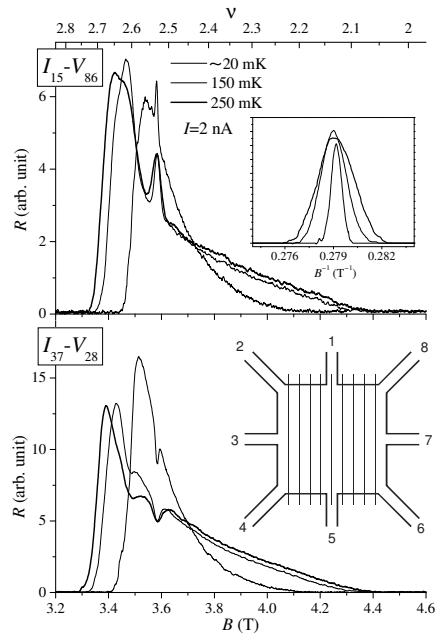}%
\caption{Magnetoresistance traces between $\nu$=2 and 3 for three temperatures. $n_e$=2.20$\times$10$^{15}$ m$^{-2}$. The direction of current is along (top) or across (bottom) the grating. Upper inset: peaks plotted against $B^{-1}$ after background subtraction. Bottom inset: schematic illustration of the sample. \label{TD}}
\end{figure}

The period $a_{\mathrm{CDW}}$ of CDW is theoretically predicted to be about 4$-$8 times the magnetic length $l=\sqrt{\hbar/eB_\perp}$ depending on the Landau index $N$.\cite{Fogler96,Koulakov96,Jungwirth99,Stanescu00} In the magnetic field range of interest, $a_{\mathrm{CDW}}$$\simeq$ 30$-$150 nm. If a 2DEG has a tendency toward spontaneously forming CDW with a period $a_{\mathrm{CDW}}$, it will undoubtedly show strong response to the external modulation having a period close to $a_{\mathrm{CDW}}$.\cite{Rezayi99} Along the same line of thought, we investigated the behavior around $\nu$=9/2 of unidirectional lateral superlattices (LSL) with periods $a$=92 and 115 nm.\cite{Endo01n} We identified small anisotropic features in magnetoresistance traces, which possibly reflect the response of CDW to the modulation. However the observed features were too small to be decisive. In the present paper, we focus on the second LL\@. We report qualitatively similar, but much more intense, anisotropic features at $\nu$=5/2 (and 7/2) for $a$=92 nm LSL (see, e.g., Fig.\ \ref{TD}). We have also observed sharp peaks at higher LL's up to $\nu$=25/2 for current along the grating (qualitatively similar as regard the direction of the current and grating), which will be reported in detail elsewhere. In the case of the $N=1$ LL, the expected role of external modulation is two-fold: to make the anisotropic state energetically favorable than the isotropic state \cite{noeven} --- the role played by $B_{\|}$ in the case of ultrahigh mobility plain 2DEG ---, and to assist the CDW to form and/or align itself, which would otherwise be obstructed by the impurities in moderate-mobility 2DEG. The observation of anisotropic transport features suggests that the external modulation \emph{do} play the expected roles. The observed peaks/dips are much more distinct than before \cite{Endo01n} due presumably to optimization of the ratio $a/l$ and also to the expansion of energy scale by larger magnetic field. Two new attempts are made in the present study: (1) samples with square geometry are used instead of Hall bars and (2) the ratio $a/l$=$a\sqrt{2\pi n_e/\nu}$ for a given $\nu$ is tuned by varying the electron density $n_e$ through infrared LED illumination. Square geometry has an advantage of allowing measurement of anisotropy within a \emph{single} LSL sample. This is quite important since it is nearly impossible to prepare two exactly identical Hall bars differing only in the orientation of the grating;\cite{Endo01n} the smallest disparity in $n_e$ can deteriorate the strict comparison. To achieve maximum response, it is desirable to make $a$ as close to $a_{\mathrm{CDW}}$ as possible. Since $a$ is fixed once a sample is prepared, we tuned $l$=$\sqrt{\nu/2\pi n_e}$ instead, to which $a_{\mathrm{CDW}}$ is predicted to be related.

\begin{figure}[t]
\includegraphics[bbllx=20,bblly=275,bburx=700,bbury=800,width=8.5cm]{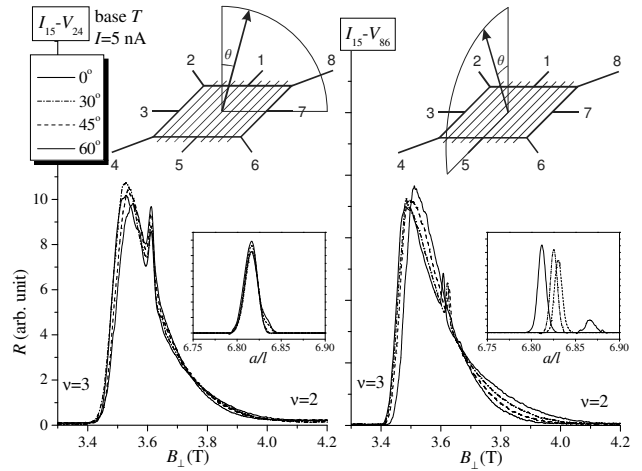}%
\caption{Magnetoresistance traces between $\nu$=2 and 3 for various tilt angles at the base temperature. $n_e$=2.20$\times$10$^{15}$ m$^{-2}$. Current is along the grating. Field is tilted toward the direction perpendicular (left) or parallel (right) to the grating. Insets show peaks plotted against $a/l$=$a\sqrt{eB_{\perp}/\hbar}$ after background subtraction. \label{RD}}
\end{figure}

The lower inset of Fig.\ \ref{TD} depicts the schematic of the sample. A square mesa (40$\times$40 $\mu$m$^2$) with eight arms (width 4 $\mu$m) is lithographically defined out of conventional GaAs/AlGaAs single heterostructure 2DEG ($\mu$$\simeq$75 m$^2$/Vs and $n_e$$\simeq$1.97$\times$10$^{15}$ m$^{-2}$ before illumination). A grating ($a$=92 nm) of electron-beam resist was placed on the square to introduce potential modulation through strain-induced piezoelectric effect as before.\cite{Endo00e,Endo01n,Endo01c} The modulation amplitude is estimated from the low-field commensurability magnetoresistance oscillation to be $\sim$0.015 meV,\cite{Endo01c} which is quite small but still much larger than the native anisotropy energy of $\sim$1 mK per electron estimated in ultrahigh mobility plain 2DEG.\cite{Cooper01} It can readily be seen that $I_{15}$-$V_{24}$ or $I_{15}$-$V_{86}$ ($I_{37}$-$V_{28}$ or $I_{37}$-$V_{46}$) mainly measures resistivity component parallel (perpendicular) to the grating, where $I_{ij}$-$V_{kl}$ denotes the probe configuration using arms $i,j$ as source/drain and $k,l$ as voltage probes. It has been pointed out that the square geometry exaggerates the anisotropy due to current-path effect.\cite{Simon99,Endo02s} However, since it is not our purpose to quantify the anisotropy, the property is rather advantageous bringing small anisotropy into light. The grating is placed with their stripes parallel either to $\langle$110$\rangle$ or to $\langle$1$\overline{1}$0$\rangle$ axis in order to maximize the piezoelectric effect.\cite{Skuras97} Although we mainly discuss the former arrangement in the following, both directions of the grating give the consistent results \cite{noteCryst} confirming that the transport is ruled by the external modulation, and the crystallographic axes do not play major role.

The main panels of Fig.\ \ref{TD} show magnetoresistance traces between $\nu$=2 and 3 at three different temperatures from our base temperature ($\sim$20 mK) up to 250 mK\@. The traces are taken after slight illumination ($n_e$=2.20$\times$10$^{15}$ m$^{-2}$). Near $\nu$=5/2, a sharp peak is observed for current parallel to the grating, which is replaced by a dip for the perpendicular current. The peak/dip broadens with temperature almost symmetrically in $B^{-1}$ hence in $\nu$ (see the upper inset), but their height/depth does not change very much in the measured temperature range. The measurements were done by ordinary ac (13 Hz) lock-in technique. The peak/dip was rather insensitive also to the measurement current in the range $I$=1--20 nA, although peak broadening attributable to the heating was seen for larger currents. Measurements with $I$=2 or 5 nA are shown throughout this paper where broadening was almost negligible. The temperature dependence is quite dissimilar from that of ultrahigh mobility plain 2DEG: in the latter, the peak height strongly depends on the temperature but the width is almost temperature insensitive, and the anisotropy vanishes at the temperature as low as 150 mK.\cite{Lilly99,Du99} Moreover, the width of our peak/dip is much narrower ($\Delta \nu_{\mathrm{FWHM}}$$\simeq$0.007 at the lowest temperature and $\simeq$0.024 even at 250 mK), accounting for only small fraction of the region between two successive integer QHE, while in ultrahigh mobility plain 2DEG, the peak spans major part of the region ($\Delta \nu_{\mathrm{FWHM}}$$\simeq$0.3--0.4). In a nutshell, our anisotropy occurs only when much severer condition for $\nu$ (or $a/l$) is met, but once it takes place, it is much more robust. The robustness of the observed anisotropic features against temperature and also against impurities does not in itself defy the interpretation by CDW\@. We believe low temperature and extremely high mobility are required for plain 2DEGs in order for the very small built-in anisotropy of unknown origin to be operative. In fact, theories predict rather high ($\agt$ 1 K) onset temperature of the anisotropic transport (with sufficient anisotropy energy), \cite{Jungwirth99} or melting temperature for CDW even in the presence of rather large disorder. \cite{Stanescu00}

The first step toward interpreting the observed transport anisotropy in terms of unidirectional CDW or stripe would be to specify the orientation of the stripe. However, this turns out to be not straightforward. Intuitively one would expect (i) the stripe aligns with the external modulation, and (ii) the low resistivity axis is along the stripe. This implies a dip for current along the grating in contradiction to the observation, suggesting that either (i) or (ii) is incorrect. Recent calculations \cite{Ishikawa01,Yoshioka01,Aoyama01} show that for external modulation with a period much larger than $a_{\mathrm{CDW}}$, the stripe tends to orient itself orthogonal to the external modulation. This counterintuitive prediction challenges the firmness of (i). We believe, however, the period $a$ of our modulation is close to $a_{\mathrm{CDW}}$. The expectation (ii) is generally believed to be valid in the interpretation of experimental data on ultrahigh mobility plain 2DEG\@. However, it is based on the assumption that the CDW is pinned by impurities and that the current is mainly carried by the stripe edge,\cite{MacDonald00,Oppen00} which probably requires to be examined more carefully. In addition, the situation may be different between our LSL and ultrahigh mobility plain 2DEG\@.

To gain more insight into the orientation of the stripe, the behavior of the peak/dip under tilted magnetic fields is investigated. Fig.\ \ref{RD} shows traces for current along the grating. By the tilt, $B_{\|}$ is introduced either perpendicular (left) or parallel (right) to the grating. As highlighted in the insets, the effect of $B_{\|}$ perpendicular to the grating is small, slightly enhancing the peak height, while the parallel $B_{\|}$ profoundly affect the peak, almost destroys the peak by $\theta$=60$^{\circ}$, shifting it to higher $B_{\perp}$. For current across the grating (not shown), basically the same trend is observed with a peak replaced by a dip. The effect of $B_{\|}$ on the stripe is theoretically calculated to be quite subtle:\cite{Jungwirth99,Stanescu00} for 2DEG with small enough thickness ($\alt 6$ nm), the stripe prefers to be oriented perpendicular to $B_{\|}$, but the trend is reversed for thicker 2DEG\@. The theories are consistent with experiments on ultrahigh mobility plain 2DEG, if one assumes the aforementioned (ii).\cite{Pan99,Lilly99t,Pan00} Our 2DEG is expected to be thinner than ultrahigh mobility 2DEG owing to higher concentration of unintentionally doped acceptors in the GaAs channel of our triangular confinement potential. In fact, a self-consistent calculation for $n_e$=2.2$\times$10$^{15}$ m$^{-2}$ and estimated acceptor concentration $N_A$=1.5$\times$10$^{20}$ m$^{-3}$ shows rms thickness of 4.8 nm. This means, according to the theories, our stripe prefers to be oriented perpendicular to $B_{\|}$, which in turn suggests, in conjunction with Fig.\ \ref{RD}, (i) mentioned above is valid and (ii) must be discarded. Considering the subtleness of thickness dependence of the calculated anisotropy energy, apparently more pieces of information are necessary to be more decisive.

\begin{figure}[t]
\includegraphics[bbllx=20,bblly=410,bburx=720,bbury=827,width=8.5cm]{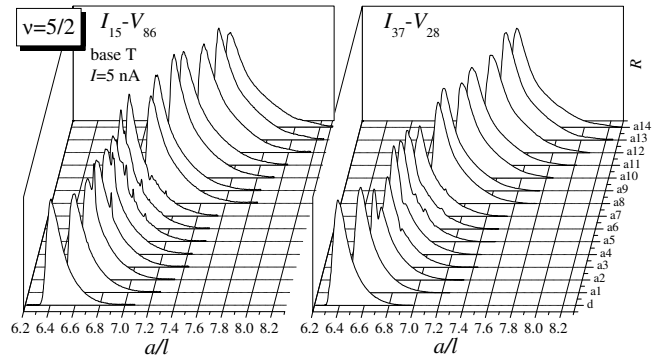}%
\caption{Evolution with $n_e$ of the region between $\nu$=2 and 3 plotted against $a/l$. Current is along (left) or across (right) the grating. $n_e$'s are increased step by step by successive LED illumination, and are (in 10$^{15}$ m$^{-2}$) d: 1.97, a1: 2.11, a2: 2.16, a3: 2.20, a4: 2.22, a5: 2.24, a6: 2.26, a7: 2.29, a8: 2.42, a9: 2.44, a10: 2.55, a11: 2.60, a12: 2.71, a13: 2.77, a14: 2.84, determined from low-field Hall resistance.\cite{accuracy} \label{fiveND}}
\end{figure}

As mentioned earlier, LED illumination is employed to vary $n_e$ and hence the ratio $a/l$. $n_e$ is varied step by step over the range 1.97--2.84$\times$10$^{15}$ m$^{-2}$.\cite{amplitude} Magnetoresistance traces between $\nu$=2 and 3 are displayed in Fig.\ \ref{fiveND}, for both current directions. The anisotropic features appear in a narrow $n_e$ range: they are observed only in traces from a1 to a7 ($n_e$=2.11--2.29$\times$10$^{15}$ m$^{-2}$), and traces d and a8 to a14 are featureless. In trace a1 very small structures begin to emerge, followed by the most conspicuous peak/dip at $a/l$$\simeq$6.7 in trace a2. The peak/dip shifts to higher field ($a/l$$\simeq$6.8) and at the same time becomes smaller in trace a3, and starts to split preserving its position at trace a4, splits into several peaks/dips and becomes complicated in traces a5 and a6. Finally in trace a7, the dip and peak interchange their positions: a dip (peak) appears in the trace with current along (across) the grating at $a/l$$\simeq$7.0.\cite{reproduce} In terms of $\nu$, the peaks/dips move in a non-monotonic way around $\nu$=5/2$\pm$0.1.\cite{accuracy}

The observation that the peak/dip is most prominent at $a/l$$\simeq$6.7 suggests that 6.7$l$ represents a certain optimum length scale of the 2DEG\@. The simplest idea is to identify 6.7$l$ with $a_{\mathrm{CDW}}$. However, theoretically estimated $a_{\mathrm{CDW}}$$=$4.443$l$ for $N$=1 LL of zero-thickness 2DEG \cite{Jungwirth99,Laguerre} is much smaller. Inclusion of the effect of finite thickness will presumably make the estimate larger by softening the short range repulsive force, but at present we are not sure whether the present discrepancy is reconciled with this inclusion. \cite{YoshiokaPC} Since 6.7$l$ is very close to 1.5$\times$4.443$l$=6.66$l$, another possibility arises that the observed ``resonance'' signifies $a$=1.5$a_{\mathrm{CDW}}$. In that condition, external modulation provides 2DEG with wave vector commensurate to the CDW and at the same time every other minimum in the potential modulation experiences the maximum and the minimum of charge density, respectively, which will make translational motion of CDW easier. Since, as mentioned earlier, our tilted-field experiment favors the alignment of the stripe parallel to the grating, sliding motion of the stripe is consistent with a resistivity minimum for the current across the grating. The insensitiveness to the measurement current suggests the lack in our case of the nonlinear I-V characteristics observed in ultrahigh mobility plain 2DEGs, \cite{Lilly99} which is also consistent with the picture that the CDW is not pinned. However, this sliding motion picture is no more than a crude speculation at present. Away from the ``resonant'' condition, the peak/dip probably try to survive making compromise among the optimum conditions for $a/l$, $\nu$ and so forth. On an occasion when multiple conditions come closer in energy together, splitting may result. The interchange in trace a7 of the peak and dip implies, if one assumes one-to-one correspondence between appearance of peak/dip and the orientation of the stripe, the stripe has turned its direction around. A recent theory \cite{Aoyama01,Zhu02} suggests the possibility of such turnaround. Since the orientation of the stripe is expected to be reflected in its response to $B_{\|}$, we have conducted tilt experiment also for a7. For $B_{\|}$ parallel to the grating (to be compared with the right panel in Fig.\ \ref{RD}), the dip/peak is observed to be more robust, consistent with the turnaround. Owing to the smallness of the observed features that makes quantitative evaluation difficult, however, it seems to be going too far to take this as evidence of the turnaround.

Around $\nu$=7/2, similar peaks/dips are observed, although with much less distinctness, over the entire range of $n_e$ available in the present experimental setup. The behavior is rather complicated to allow simple analysis.

To summarize, we have observed anisotropic features around $\nu$=5/2 (and 7/2) in 2DEG subjected to weak external modulation having a period comparable to the theoretically calculated period $a_{\mathrm{CDW}}$. The advantage of the present study is that the source of the anisotropy is well-defined. Nevertheless the interpretation of the transport anisotropy in terms of CDW requires further study.

\begin{acknowledgments}
The authors acknowledge D.\ Yoshioka for discussion on calculated $a_\mathrm{CDW}$ and the effect of finite thickness. This work was supported in part by a Grant-in-Aid for Encouragement of Young Scientist (13740177) and for COE Research (\#12CE2004
``Control of Electronics by Quantum Dot Structures and
Its Application to Advanced Electronics'')
from
the Ministry of Education, Culture, Sports, Science and Technology.
\end{acknowledgments}

\bibliography{fivehlvs,ninehlvs,ourpps,lsls,notefh}

\begin{thebibliography}{43}
\bibitem[{noe()}]{noeven}
\bibinfo{note}{The mobility of our present 2DEG wafer was not enough for
  even-denominator FQHE's to develop even without the modulation.}
\bibitem[{not()}]{noteCryst}
\bibinfo{note}{The latter arrangement showed slightly smaller structures, which
  we attribute to slightly worse quality of the LSL\@.}
\bibitem[{acc()}]{accuracy}
\bibinfo{note}{The accuracy of $n_e$, hence of $\nu$, is limited by hysteresis
  of the superconducting magnet at low field. For comparison, all the $n_e$'s
  are calculated using Hall resistance measured in the same sweep direction and
  rate. A systematic error in the absolute value does not affect the
  discussion.}
\bibitem[{amp()}]{amplitude}
\bibinfo{note}{Concomitantly $\mu$ varies from 75 to 110 m$^2$/Vs. The
  amplitude of potential modulation also varies due to changes in screening.
  However, we estimate the effect to be small in this relatively small $n_e$
  range.}
\bibitem[{rep()}]{reproduce}
\bibinfo{note}{Some of the features mentioned here are very small. However they
  reproduce among several sweeps including up-sweeps and down-sweeps for both
  $I$=2 and 5 nA and well distinguishable from random noises.}
\end{thebibliography}

\end{document}